\documentclass{aa}  

\makeatletter
\renewcommand*\aa@pageof{, page \thepage{} of \pageref*{LastPage}}
\makeatother

\usepackage{siunitx}
\usepackage{booktabs}

\usepackage{upgreek}
\usepackage{graphicx}
\usepackage{txfonts,textcomp}
\usepackage[colorlinks=true, linkcolor = red, citecolor=blue, urlcolor=magenta]{hyperref}

\begin{document}

   \title{The impact of stellar bars on star-formation  quenching: Insights from a spatially resolved analysis in the local Universe}
   \titlerunning{Spatially resolved indications of bar quenching}
    
   \subtitle{}

   \author{Letizia Scaloni
          \inst{1,2,}\thanks{\email{letizia.scaloni@inaf.it}}
          \and 
          Giulia Rodighiero
          \inst{3,4}
          \and
          Andrea Enia
          \inst{1,2}
          \and 
          Carlotta Gruppioni
          \inst{2}
          \and
          Francesca Annibali
          \inst{2}
          \and
          Laura Bisigello
          \inst{5,3}
          \and
          Paolo Cassata
          \inst{3,4}
          \and
          Enrico Maria Corsini
          \inst{3,4}
          \and
          Viviana Casasola
          \inst{5}
          \and
          Cristina Maria Lofaro
          \inst{6,3}
          \and
          Alessandro Bianchetti
          \inst{3,4}
          }

   \institute{Department of Physics and Astronomy ``Augusto Righi'', University of Bologna,
              Via Piero Gobetti 93/2, I-40129 Bologna, Italy
              \and
              INAF -- Astrophysics and Space Science Observatory of Bologna, Via Piero Gobetti 93/3, I-40129 Bologna, Italy 
              \and
              Department of Physics and Astronomy ``Galileo Galilei'', University of Padova, Vicolo dell'Osservatorio 3, I-35122 Padova, Italy
              \and
              INAF -- Astronomical Observatory of Padova, Vicolo dell'Osservatorio 5, I-35122 Padova, Italy
              \and
              INAF -- Radioastronomy Institute of Bologna, Via Piero Gobetti 101, I-40129 Bologna, Italy 
              \and
              Institute of Astrophysics, Foundation for Research and Technology -- Hellas (FORTH), GR-70013 Heraklion, Greece\\
              }

   \date{Received XXX; accepted XXX}

 
  \abstract
   {Stellar bars are common morphological structures in the local Universe; according to optical and NIR surveys, they are present in about two-thirds of disc galaxies. These elongated structures are also believed to play a crucial role in secular evolutionary processes, because they are able to efficiently redistribute gas, stars, and angular momentum within their hosts, although it remains unclear as to whether they enhance or suppress star formation. A useful tool to investigate this ambiguity is the main sequence (MS) relation, which tightly links stellar mass ($M_{\star}$) and star formation rate (SFR).}
   {The main goal of this work is to explore star-formation processes in barred galaxies in order to assess the relevance of bars in star-formation quenching and whether or not they affect the typical log-linear trend of the resolved MS.}
   {To this purpose, we carried out a spatially resolved analysis on subkiloparsec (subkpc) scales for a sample of six nearby barred galaxies. We collected multi-wavelength photometric data from far-ultraviolet (FUV) to far-infrared (FIR) from the DustPedia database and applied a panchromatic spectral energy distribution (SED) fitting procedure on square apertures of fixed angular size (\SI{8}{\arcsec} $\times$ \SI{8}{\arcsec}) using the \texttt{magphys} code.}
   {For each galaxy, we obtain the distributions of stellar mass and SFR surface density and relate them in the $\log \Sigma_{\star}$ -- $\log \Sigma_{\rm SFR}$ plane, deriving the spatially resolved MS relation. Although significant galaxy-to-galaxy variations are in place, we infer the presence of a common anti-correlation track in correspondence with the bar-hosting region, which shows systematically lower SFRs. This central quiescent signature can be interpreted as the result of a bar-driven depletion of gas reservoirs and a consequent halting of star formation. Our findings appear to support an inside-out quenching scenario.} 
   {} 


   \keywords{galaxies: evolution --
             galaxies: spiral --
             galaxies: star formation --
             galaxies: stellar content --
             galaxies: structure
               }

   \maketitle



\section{Introduction}

Internal processes typically dominate the late-stage evolution of galaxies, when gravitational interactions and mergers become less frequent as the Universe expands \citep{Kormendy2004}. Stellar bars are among the main internal drivers of the secular evolution of disc galaxies, because they are able to shape the current properties of their hosts, redistributing the gas, stars, and angular momentum within them \citep[e.g.][]{Athanassoula2002, Sheth2005, Debattista2006, Athanassoula2013, DiMatteo2013, Sellwood2014, Romeo2023}. 

The pivotal role of bars in improving our understanding of galaxy evolution is also demonstrated by the fact that they are common morphological structures in the local Universe. According to optical and near-infrared (NIR) surveys, between $\sim40\%$ and $\sim70\%$ of nearby disc galaxies host stellar bars \citep[e.g.][]{Eskridge2000, Barazza2008, Aguerri2009, NairAbraham2010, Buta2015, Erwin2018}. 
The evolution of this fraction over cosmic time may shed light on the epoch of bar formation. In particular, as found by several observational studies, there seems to be a consensus on a decline in the barred galaxy fraction from the local Universe up to $z \sim 1$ by at least a factor of two \citep[e.g.][]{Sheth2008, Melvin2014}. More recently, the unprecedented sensitivity and spatial resolution in the NIR of the James Webb Space Telescope (JWST) allowed authors to infer the evolution of the bar fraction at higher redshifts. Prominent elongated structures have been identified in disc galaxies up to $ z \sim 3.8$ \citep[e.g.][]{Guo2023, Costantin2023, Amvrosiadis2024}, while \cite{LeConte2023} presented a full census of observable stellar bars at $z > 1$. These latter authors pointed out that the incidence of bars in disc galaxies decreases from $\sim19\%$ to $\sim7\%$ going from $1 \leq z \leq 2$ to $2 < z \leq 3$. Consistent results were also obtained in a fully cosmological context thanks to a new generation of high-resolution simulations \citep[e.g.][]{Fragkoudi2020, Zhao2020, RosasGuevara2022}.

The fact that both observational studies and simulations present evidence of a significant population of barred galaxies up to $z \sim 2$ suggests that stellar bars are already in place in the early Universe, highlighting the possibility that bar-driven secular processes influence galaxy evolution over billions of years. 
Bars are thought to form over timescales of the order of a hundred million years, once a dynamically cold and rotationally supported disc has settled \citep[e.g.][]{Ostriker1973, Sellwood1993}. The onset of bar formation occurs mainly when the stellar orbits of massive disc galaxies deviate from a circular path due to instabilities. 

Whether the growth of stellar bars enhances or suppresses star formation is still debated \citep[e.g.][]{Kim2017, Wang2020}. However, there are indications supporting the hypothesis that bars can suppress star formation (bar quenching) in a region spanning from the central subkiloparsec (subkpc) to the end of the bar itself \citep[e.g.][]{Masters2010, Masters2012, Gavazzi2015, James2016, George2019a, RosasGuevara2020, Zhang2021}, thus contributing to the observed transition from active star-forming galaxies to passive ones \citep[e.g.][]{ManBelli2018}. This could be due to the bar-induced shock and shear; these stabilise the gas against collapse by increasing turbulence, which, in turn, results in the inhibition of star formation \citep[e.g.][]{Haywood2016, Khoperskov2018}. Another possible mechanism is related to the redistribution of cold gas along the length of the stellar bar. The torque produced by this morphological structure perturbs the interstellar medium (ISM), driving gas and dust inflow towards the galactic centre \citep[e.g.][]{Casasola2011, Combes2013, Spinoso2017}. As a consequence, this boosts the star formation in the inner subkpc-scale area and depletes the bar-hosting region of fuel for further star formation \citep[e.g.][]{George2019a, George2020, Newnham2020}. 

The bar-quenching scenario suggested by many recent observations and simulations predicts that bars can be responsible for the onset of a phase of high star formation rate (SFR) at their early evolutionary stages, potentially triggering starburst events or fuelling an active galactic nucleus (AGN), accompanied by a subsequent rapid consumption of gas, thus inducing a premature quenching phase in their host galaxies \citep[e.g.][] {Spinoso2017, FraserMcKelvie2020a, FraserMcKelvie2020b, George2021, Geron2021}. 
This could explain why some literature works agree on the fact that barred galaxies have higher SFRs than unbarred ones, especially at the centre, along the length of the bar, and within the circumnuclear region \citep[e.g.][]{Sheth2005, Zurita2008, Ellison2011, Coelho2011, Prieto2019, Lin2020, SanchezGarcia2023}. Conversely, many other studies have pointed out that bars are most likely to be found in quiescent galaxies, which are typically more massive, redder, and gas-poor \citep[e.g.][]{Vera2016, CervantesSodi2017, Geron2021}.

In this context, it is important to further explore the impact of bar quenching by studying the star-formation properties of barred galaxies and their dependence on the presence of the bar, especially through the existing empirical correlations between the main structural parameters of galaxies. In particular, the stellar mass ($M_{\star}$)--SFR relation is one of the most useful tools in modern astrophysics \citep[e.g.][]{Brinchmann2004, Noeske2007_2, Noeske2007_1, Salim2007}. 
Habitually referred to as the main sequence (MS) of star-forming galaxies (SFGs), it shows that more massive galaxies host larger amounts of star formation than those of  lower mass. Many literature works suggest that this log-linear correlation is valid ---with an intrinsic scatter of $\sigma$ $\sim$ 0.3 dex--- for both local and distant galaxies (up to z $\sim$ 6) and for a wide range of stellar masses and considering different SFR tracers \citep[e.g.][]{Daddi2007, Pannella2009, Santini2009, Oliver2010, Elbaz2011, Lee2012, Whitaker2012, Moustakas2013, Whitaker2014, Tasca2015, Barro2017, Santini2017, Pearson2018, Popesso2019b, Daddi2022, Popesso2023}. 
Galaxies exhibiting various rates of stellar production (starbursts, SFGs, and passive galaxies) populate different regions of the $M_{\star}$--SFR plane. Therefore, their position on the MS can be used as a diagnostic tool to identify their state of star formation and to figure out the origin of their colour bimodality \citep[e.g.][]{Rodighiero2011, Renzini2015, Bisigello2018, Popesso2019a}.

A significant limitation in our understanding of the $M_{\star}$--SFR relation stems from the fact that many studies have relied on the integrated properties of galaxies, without distinguishing among their morphological components. A more advanced approach involves characterising this correlation on smaller scales employing spatially resolved data. Thanks to high-spatial-resolution imaging and the advent of integral field spectroscopy (IFS) facilities, the MS relation was found to also be valid locally, at kpc and subkpc scales \citep[e.g.][]{Sanchez2013, Wuyts2013, CanoDiaz2016, Abdurro'uf2017, Hsieh2017, Lin2019, Enia2020, Ellison2021, Casasola2022}. This implies that the star-formation activity originates from physical mechanisms that could be universal across various physical scales.
There is a general consensus on the existence of such a `resolved' MS relation between the locally measured SFR surface density ($\Sigma_{\rm SFR}$) and the stellar mass surface density ($\Sigma_{\star}$). However, the slope, zero-point, and scatter of this relation are subject to significant galaxy-to-galaxy variation, and depend on the different morphological structures, SFR tracers, dust-extinction correction, and fitting procedures \citep[e.g.][]{Abdurro'uf2017, Vulcani2019, Pessa2021, Pessa2022, Casasola2022}. 

With the aim of probing the relevance of bar-driven quenching, several observational works and simulations analysed the integrated MS relation of statistically large samples of barred galaxies \citep[e.g.][]{Vera2016, George2019b, FraserMcKelvie2020a, George2021, RosasGuevara2020, RosasGuevara2022}. However, to the best of our knowledge, there are just a few studies that investigated the spatially resolved MS across different galactic structures, including the bar \citep{Pessa2021, Pessa2022}. Moreover, previous literature works are almost all based on spectroscopic data or UV-to-optical tracers of SFR \citep[see][for a review]{Sanchez2020}.

In the present work, we further investigated the MS relation at subkpc scales for six barred galaxies in the local Universe, exploiting the power of multi-wavelength photometric data ---from far-ultraviolet (FUV) to far-infrared (FIR)--- collected in the DustPedia archive \citep{Davies2017, Clark2018} and building on the methodology proposed by \cite{Enia2020}. With this approach, we retrieve a reliable measure of the SFR, taking into account both the unobscured and the dust-obscured components, once a spectral energy distribution (SED) fitting procedure is applied. The final aim of this study is to get a more comprehensive view of the star-formation processes in barred galaxies and to understand whether the presence of the bar affects the typical log-linear trend of the resolved MS, at least in the inner regions.

The paper is structured as follows: in Section\,\ref{sec:data_sample}, we briefly describe the data used in this work and our sample selection; in Section\,\ref{sec:data_analysis}, we report the main steps of the methodology used to derive $M_{\star}$ and SFR; in Section\,\ref{sec:results}, we present our results concerning the spatially resolved MS; in Section\,\ref{sec:discussion}, we discuss the implications of our findings. Finally, in Section\,\ref{sec:conclusions}, we summarise our work and the main outcomes. Throughout the paper, we assume a $\Lambda$CDM cosmology with parameters $\Omega_{m} = 0.308$, $\Omega_{\Lambda} = 0.692$, and $H_0 = 67.8\,{\rm km\,s^{-1}\,Mpc^{-1}}$ \citep{PlanckCollaboration2016}, and a \cite{Chabrier_2003} initial mass function (IMF).


\section{Data and sample}
\label{sec:data_sample}

\subsection{DustPedia archive}
DustPedia\footnote{The DustPedia website is available at \url{http://dustpedia.astro.noa.gr/}} is a research project that was developed to enable a comprehensive characterisation of the cosmic dust in the local Universe and of its effects on multi-wavelength observations of ISM in galaxies \citep{Davies2017}. The DustPedia sample consists of 875 galaxies that were observed by \textit{Herschel} \citep{Pilbratt2010} and that lie within $\sim$ 50 Mpc, with an optical diameter D$_{25}$\footnote{D$_{25}$ is the major axis of the 25-th level isophote, which is the one at which the value of the surface brightness falls below 25mag/arcsec$^2$ in the B band.} of greater than \SI{1}{\arcmin}. The complete data set, which is publicly available to the scientific community, contains standardised multi-wavelength imagery and photometry of the whole sample of galaxies, including both custom reductions and archival data, spanning over five orders of magnitude in wavelength from UV to microwave \citep{Clark2018}.

\begin{table*}
        \centering
        \caption{Main properties of the galaxy sample.}\label{tab:properties}
        \begin{tabular}{ccccccccccc}
            \toprule
                \toprule
                Galaxy & RA & Dec & $D$ & $i$ & R$_{25}$ & R$_{\rm bar}$ & $\log(M_{\rm \star} $/$\si{M_\odot}$)  & SFR & Cell size & Hubble \\
                Name & [deg] & [deg] & [Mpc] &[\textdegree] & [kpc]& [kpc] &[dex] & [$\si{M_\odot}\,{\rm yr}^{-1}$] & \SI{8}{\arcsec} [kpc] & Type\\
                \midrule        
                NGC~1566 & $\phantom{0}65.00160$ & $-54.93781$ & $\phantom{0}6.14$ & 47.9 & $\phantom{0}6.31$ & 1.2 & $\phantom{0}9.57^{+0.15}_{-0.23}$ & $0.44 \pm 0.08$ & 0.24 & SABb\\ 
            \noalign{\vskip 3pt}
                NGC~3351 & $160.99035$ & $+11.70355$ & $\phantom{0}9.91$ & $54.6$ &10.17 & 3.8 & $10.10^{+0.09}_{-0.11}$ & $0.68 \pm 0.15$ & 0.38 & SBb\\
            \noalign{\vskip 3pt}
                NGC~3953 & $178.45425$ & $+52.32653$ & $12.47$ & 62.1 & $10.89$ & 2.8 & $10.13^{+0.11}_{-0.14}$ & $0.36 \pm 0.29$ & 0.48 & SBbc\\ 
            \noalign{\vskip 3pt}
                NGC~4579 & $189.43140$ & $+11.81800$ & $19.95$ & 41.9 & $14.14$ & 4.3 & $10.88^{+0.05}_{-0.06}$ & $0.57 \pm 0.29$ & 0.77 & SBb \\ 
            \noalign{\vskip 3pt}
                NGC~4725 & $192.61095$ & $+25.50091$ & $12.08$ & 45.4 & $16.72$ & 7.0 & $10.64^{+0.07}_{-0.08}$ & $0.58 \pm 0.07$ & 0.47 & SABa \\
            \noalign{\vskip 3pt}
                NGC~5236 & $204.25365$ & $-29.86556$ & $\phantom{0}4.90$ & $14.1$ & $\phantom{0}9.38$ &  2.4 & $10.27^{+0.10}_{-0.13}$ & $4.20 \pm 0.69$ & 0.19 & SBc \\           
                \bottomrule
        \end{tabular}
        \tablefoot{Galaxy name, coordinates in J2000 system reference, distances, inclinations, R$_{25}$ sizes and morphological classifications (Hubble type) are the same as those adopted by the DustPedia collaboration, and come from the HyperLEDA database. R$_{\rm bar}$ sizes are taken from \cite{MunozMateos2013} and \cite{Salo2015}. The values of $M_{\star}$ (in logarithmic scale) and SFR are obtained from \cite{Nersesian2019}. Cell sizes refer to the pixel-by-pixel SED fitting procedure (in this case, with an \SI{8}{\arcsec} square size reported in kpc).}
\end{table*}

\subsection{Sample selection}
To properly select a sample of barred spiral galaxies from the DustPedia database, we first put some constraints on the Hubble stage index T\footnote{The Hubble stage index T is a numerical value, running from -6 to 10, that is assigned to each morphological class of galaxies in the De Vaucouleurs scheme. Negative values correspond to early-type galaxies (ellipticals and lenticulars), while positive numbers correspond to late-type galaxies (spirals and irregulars).}, restricting its range between 2 and 7. This allows us to exclude early-type galaxies, intermediate classifications between lenticular and spirals, and also irregular galaxies. 
We then applied a cutoff in distance, at approximately 20 Mpc, and also imposed a lower limit on the optical diameter D$_{25}$, considering only galaxies with D$_{25} >$ \SI{4}{\arcmin}. Both these conditions ensure a selection of sufficiently close and extended galaxies, so that their central regions hosting the bar can be properly investigated with sufficient resolution. Furthermore, we focus on sources that have a disc inclination of $i <$ 65\textdegree to avoid projection corrections. In addition, we selected only spiral galaxies that show a bar structure based on their known morphological classification. We also exclude objects with peculiar structures and those hosting obvious AGN (according to the literature). Indeed, the presence of strong nuclear activity would largely affect the emission from star formation in most of the galaxy and especially in the central region, which is the area we are investigating in order to probe the relevance of the bar-driven quenching. Finally, following these outlined constraints, we obtained a sample that is as free as possible from any contamination, enabling us to more effectively distinguish the impact of the bar.

The final sample consists of six objects: NGC~1566, NGC~3351, NGC~3953, NGC~4579, NGC~4725, and NGC~5236. Distances, inclinations, sizes, and integrated properties of the galaxies under consideration are presented below and summarised in Table \ref{tab:properties}. 
All the sources are identified by their own equatorial coordinates  in right ascension (RA) and declination (Dec) at epoch J2000. For each galaxy, along with the morphological type, we report: its luminosity distance in megaparsecs (Mpc), inclination, R$_{25}$ (semi-major axis of the 25th level isophote), and size in kpc. All these quantities are taken from the HyperLEDA database \citep{Makarov2014} and are those adopted by the DustPedia collaboration. Moreover, we report the bar radius R$_{\rm bar}$ in kpc, as determined in previous studies based on photometric decompositions of images from the \textit{Spitzer} Survey of Stellar Structure in Galaxies (S$^{4}$G). In particu\-lar, we retrieve the bar radius of NGC~1566 and NGC~3351 from \cite{MunozMateos2013}, while we refer to the work of \cite{Salo2015} 
for details of the bars of the other galaxies. We then show the total stellar mass and the total SFR ---with the corresponding uncertainties--- of all the selected sources. These integrated properties are expressed in units of $\si{M_\odot}$ and $\si{M_\odot}$yr$^{-1}$,  respectively, and are taken from the results of \cite{Nersesian2019}, paying attention to properly rescale them in accordance with a \cite{Chabrier_2003} IMF. \cite{Nersesian2019} derived the global properties of most of the DustPedia sample using the aperture-matched photometry generated by the Comprehensive and Adaptable Aperture Photometry Routine (\texttt{CAAPR}), which is presented in \cite{Clark2018}. On this photometric dataset,  \cite{Nersesian2019} modelled the panchromatic SED employing the \texttt{cigale} fitting code \citep{Boquien2019} and assuming the \cite{Salpeter1955} IMF. We also indicate the size (in kpc) of the square apertures adopted for the spatially resolved analysis presented in this paper (i.e. the size of the cells). In particular, this value is related to the so-called pixel-by-pixel SED-fitting procedure that is applied to a grid of square cells of \SI{8}{\arcsec} per side, which corresponds to different physical sizes depending on the distance of the galaxy. Further details of the methodology are provided in Sect.\,\ref{sec:data_analysis}.

\subsection{Sample description}

Here, we outline the main characteristics of the selected galaxies, building on previous literature findings and highlighting those most relevant to our analysis.
Firstly, it is important to note that although we decided to rule out any possible strong AGN contamination, our sample includes two galaxies in which there are hints of activity in their nuclear regions, namely
NGC~1566 and NGC~4579.  
NGC~1566 is considered a low-luminosity AGN (LLAGN), but it shows a soft X-ray spectrum that is likely explained with a luminous nuclear starburst \citep{Tomas2022}. Thanks to ALMA observations, \cite{Combes2014} found that the HCO$^{+}$(4-3) line is three times stronger than the HCN(4-3), as expected when star formation excitation dominates over AGN heating. Also, \cite{Vermot2023} suggested that the nucleus of this galaxy hosts both phenomena: very recent or ongoing star formation and an AGN (accretion disc + hot dust). The latter contribution is supported by detection in optical and NIR spectra of associated emission lines \citep{Smajic2015, daSilva2017}. This galaxy is characterised by variable nuclear activity and observations of the AGN in the outburst phase seem to identify disc instabilities as the cause of the variability \citep{Parker2019}.
NGC~4579 was classified as an intermediate LINER/Seyfert 1.9 spiral galaxy by \cite{Ho1997}. Subsequent works, such as \cite{GarciaBurillo2009}, considered this object a representative example of a LLAGN, in which several mechanisms can co-operate to transport the gas from the outer disc to the central regions (spiral arms, bar, viscosity), thus feeding star formation and possibly the AGN activity itself. More recently, this galaxy was studied in terms of variability timescales: \cite{Younes2019} found a lack of variability in the X-ray emission and a broad Fe K$\upalpha$ line, arguing for an altered accretion geometry in LLAGN compared to luminous AGN. 

We decided to include both of these objects in our sample because their LLAGN appears to have only a marginal impact on the overall galaxy properties and on its energy budget \citep{Casasola2015}. This nuclear activity could at most affect the inner subkpc region. For example, \cite{deSaFreitas2023} masked a region of $\sim$~\SI{2}{\arcsec} at the centre of NGC~1566 in order to avoid any potential contami\-nation in their analysis, which was based on the derivation of kinematic and stellar population properties in the nuclear disc. This angular scale is smaller than our typical aperture employed in the subsequent spatially resolved procedure (\SI{8}{\arcsec}). For this reason, we initially assumed that AGN activity in their nuclear region does not affect our data analysis, and then we verified this assumption a posteriori by looking at the SED fitting in the central pixels (see Sect.\,\ref{sec:discussion}).

NGC~3351 is actively forming stars in its nucleus and in a circumnuclear ring (with a diameter of $\sim$ 0.7 kpc), as is highlighted by many multi-wavelength studies performed on subkpc scales \citep{Elmegreen1997, Bresolin2002, Ma2018}. H$\upalpha$ imaging of this galaxy shows that the bar region is devoid of emission \citep{James2009} and, according to \cite{James2016}, is characterised by an old stellar population. 
The existing literature suggests that the nuclear starburst and the observed suppression of recent star formation ($\sim 10\,{\rm Myr}$) in the bar region ---which lacks neutral and molecular hydrogen--- are likely due to the influence of the bar itself \citep[e.g.][]{George2019a}. 

On the contrary, there are very few studies concerning NGC~3953 and NGC~4725. The former has a rather low bulge/disc ratio, suggesting that its inner regions are mostly dominated by components other than the bulge, such as the bar and ring \citep{Castro-Rodriguez2003}. The work of \cite{Staudaher2019} seems to show the presence of some leftover turbulent gas in this galaxy, resulting in a newly discovered looping tidal stream. On the other hand, NGC~4725 was identified as a double-barred galaxy by \cite{Erwin2004}, but \cite{deLorenzoCaceres2013} suggested that, at present, the inner bar is playing a minor role in promoting major morphological changes. \cite{Chiaraluce2019} attempted to measure its nuclear radio emission, including this object among a sample of low-luminosity Seyfert galaxies \citep{VeronCetty2006}. However, their Very Large Array (VLA) observations produced an unresolved image and a peculiar radio spectrum, which merits further investigation to derive meaningful information from this source. 
 
Regarding NGC~5236, this galaxy has a grand-design barred spiral morphology and hosts a nuclear starburst ring \citep{Buta1993, Calzetti2004, Knapen2010} leading to a high total SFR of $\sim$ 4.2 $\si{M_\odot}$yr$^{-1}$ \citep[][]{Leroy2021}. Furthermore, \cite{DellaBruna2022} observed that both the stars and gas kinematics of this galaxy trace regular rotation along the main star-forming disc as well as a fast-rotating circumnuclear disc of \SI{30}{\arcsec} ($\sim$ 700 pc) in diameter, likely originating from secular processes driven by the galactic bar (as already reported by \cite{Gadotti2020}).  \cite{DellaBruna2022} also detected an extended ionised gas region, which could be attributed to a bar-driven inflow of shocked gas, among other factors.

\subsection{Environmental properties}
\label{subsec:environment}
The surrounding environment of galaxies is believed to play a crucial role in their evolutionary path \citep{Dressler1980, Poggianti1999}. According to many observational works, late-type galaxies (spiral, blue, and star forming) are more likely to be found in lower density environments than early-type ones (elliptical, red, and quenched), which are more numerous in groups, clusters, and interacting systems \citep[e.g.][]{Casasola2004, Blanton2005, Peng2010, Boselli2014}.
Indeed, there are several external processes that are commonly identified as responsible for this phenomenology and that can alter the effects of internal and/or secular processes, such as the action of a stellar bar. Among them, we can mention: ram-pressure stripping \citep{Gunn1972, Yun2019}, mergers \citep{Milos1994}, harassment \citep[high-speed encounters between galaxies;][]{Moore1996}, and strangulation \citep[cutoff of gas accretion;][]{Kawata2008}.   
Below, we present the environment in which each of our barred galaxies is located; we investigated this aspect  in order to search for any possible influence of environment on star formation.

NGC 1566 is the brightest member of the Dorado group \citep{Aguero2004, Kilborn2005} and belongs to a pair-like structure where its companion is the dwarf elliptical NGC~1581 \citep{Rampazzo2022}.
The work of \cite{Elagali2019} found that NGC~1566 has an asymmetric and mildly warped HI disc. To explain this feature, these authors excluded the scenario of past interactions with other galaxies in favour of the possibility of having experienced ram-pressure stripping from the intergalactic medium (IGM), at least in its outskirts; however, the high atomic gas fraction that characterises this galaxy hints at a subtle ram-pressure stripping effect, as expected in galaxy groups.

We obtained information on the environmental properties of NGC~3351 and NGC~5236 from the catalogue produced by \cite{Karachentsev2013}. This latter contains information about 869 galaxies within 11 Mpc and includes three estimates of the `tidal index', which quantifies the local density environment. In particular, we looked at the tidal index $\uptheta_{5}$, which is determined by the five most important neighbours. Positive values of this parameter indicate group membership, while negative values correspond to either isolated galaxies or galaxies in the outskirts of a group. NGC~3351 has a relatively high value of $\uptheta_{5} = 1.2$ compared to the distribution of galaxies in the \cite{Karachentsev2013}, meaning that it is in a mildly dense environment. On the other hand, NGC~5236 has $\uptheta_{5} = - 0.1$ and so is considered a `field' galaxy. In support of this definition, we mention that NGC 5236 is present in the catalogue of isolated galaxies of \cite{Bettoni2003}.  

For what concerns NGC~3953, \cite{Kourkchi2017} identified it as part of a group, whose main member is the barred spiral NGC~3992.
According to \cite{Vollmer2012}, NGC~4579 is one of the brightest galaxies in the Virgo Cluster. It is an anemic galaxy and has a relatively small amount of neutral hydrogen located inside its disc and concentrated in clumps \citep{Cayatte1990}. This deficiency of gas content is thought to be a consequence of interactions with the intracluster medium of the Virgo Cluster, suggesting that ram-pressure stripping may have occurred in the galaxy's outskirts \citep{Koopmann2004}.

NGC~4725 is the brightest member of the Coma I aggregation \citep{Gregory1977} and seems to be interacting with the neighbouring spiral galaxy NGC~4747 \citep{Haynes1979}. Gravitational interactions between these two galaxies likely resulted in a gas-rich tidal tail that extends about
\SI{8}{\arcmin} ($\sim$ 30 kpc) to the northeast of NGC~4747 and appears to
contain two distinguishable star clusters \citep{LeeWaddell2016, LeeWaddell2018}. On the other hand, only a relatively small distortion in the HI distribution is observed in NGC~4725.

\subsection{Multi-wavelength observations}
As briefly mentioned above, DustPedia provides multi-wavelength observations from UV to microwave. Regarding both Spectral and Photometric Imaging REceiver \citep[SPIRE;][]{Griffin2010} and Photodetector Array Camera and Spectrometer \citep[PACS;][]{Poglitsch2010} data, the \textit{Herschel} Science Archive (HSA\footnote{https://archives.esac.esa.int/hsa/whsa/}) was queried to find all \textit{Herschel} photometric observations covering each DustPedia galaxy. The ancillary imaging data were gathered from a further  seven facilities that observed large numbers of nearby galaxies: the GALaxy Evolution eXplorer \citep[GALEX;][]{Morrissey2007}, the Sloan Digital Sky Survey \citep[SDSS;][]{York2000, Eisenstein2011}, the Digitized Sky Survey (DSS), the 2 Micron All-Sky Survey \citep[2MASS][]{Skrutskie2006}, the Wide-field Infrared Survey Explorer \citep[WISE][]{Wright2010}, the \textit{Spitzer} Space Telescope \citep{Werner2004}, and \textit{Planck} \citep{Planck2011}. More information on the different data sets available can be found in \cite{Enia2020} and references therein.

Thanks to the data derived from these multi-wavelength observations collected in the DustPedia archive, it is possible to exploit the power of complete multi-band photometric coverage, and especially its ability to provide a measure of the comprehensive SFR in galaxies, directly taking into account both the optical and dust-obscured components.
For what concerns our work, we use observations from GALEX, SDSS, 2MASS, WISE, \textit{Spitzer,} and \textit{Herschel}, going from the FUV (1516~\AA) to the FIR (350 $\upmu$m). However, it has to be noted that not all of these data sets are accessible for the whole sample. Indeed, the entire set of 23 photometric bands is only  available for NGC~3351, NGC~4579, and NGC~4725. NGC~3953 is observed in 20 bands; GALEX filters (FUV, NUV) and PACS 70$\upmu$m photometry are not available for this object. Instead, 18 bands are accessible for NGC~1566 and NGC~5236, which lack optical observations (the five Sloan optical filters).  

\begin{figure*}
    \centering
    \includegraphics[width=\linewidth]{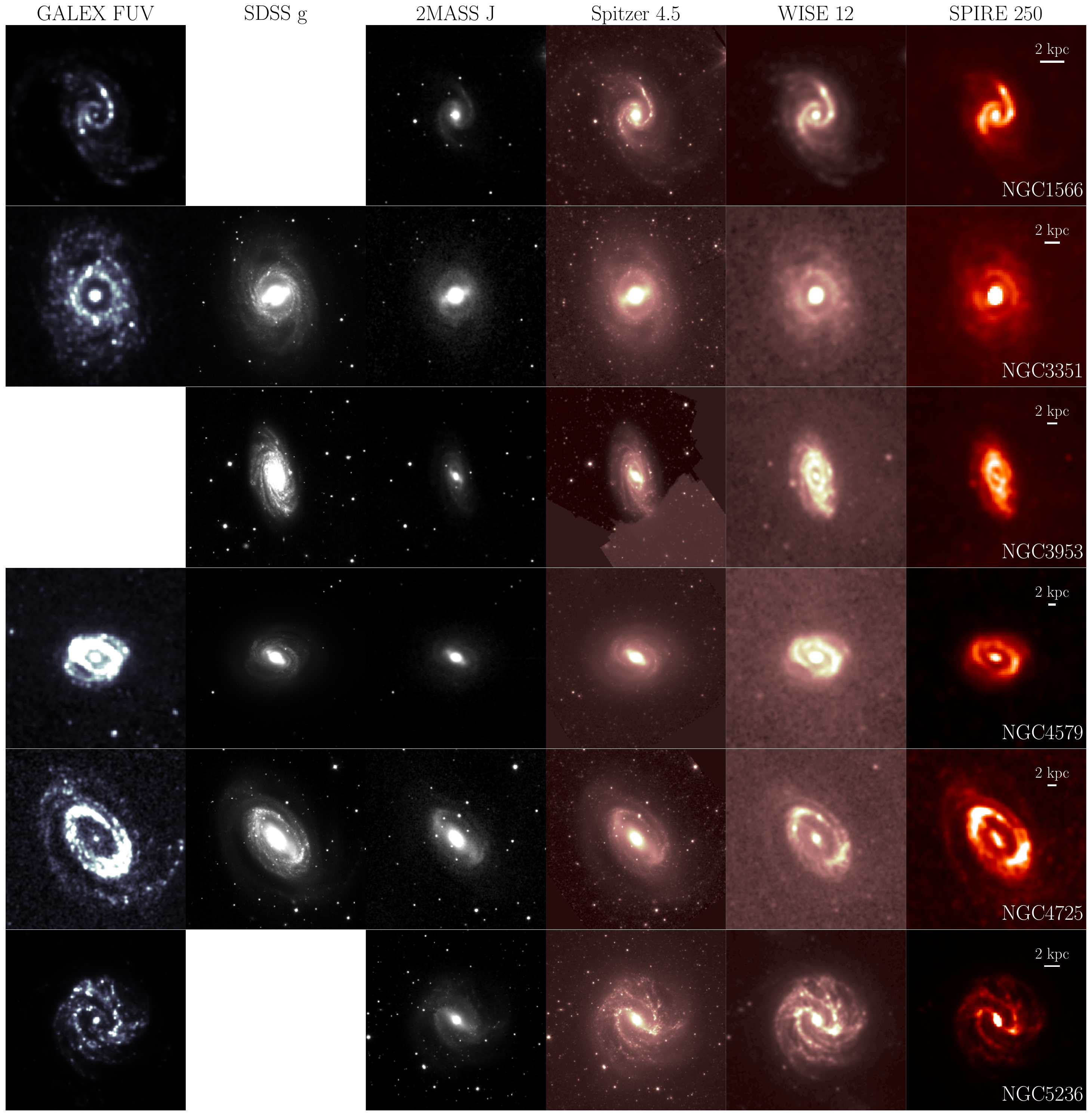}
    \caption{Example of six-band imaging available for NGC~1566, NGC~3351, NGC~3953, NGC~4579, NGC~4725, and NGC~5236 (from the DustPedia archive). Each row represents a galaxy, while in columns we display the observations in different bands: GALEX FUV, SDSS g, 2MASS J, \textit{Spitzer} 4.5 $\upmu$m, WISE 12 $\upmu$m, and SPIRE 250 $\upmu$m.} 
    \label{fig:multi-wave_example}
\end{figure*}

Figure \ref{fig:multi-wave_example}  shows the selected galaxies as they are observed in six different bands: FUV with GALEX, g-band with SDSS, J-band with 2MASS, 4.5 $\upmu$m with \textit{Spitzer}, 12 $\upmu$m with the WISE, and 250 $\upmu$m with \textit{Herschel}. An elongated morphological structure is clearly visible in all the galaxies, both in the SDSS optical bands (when available) and in the NIR ones (2MASS and \textit{Spitzer}).


\section{Data analysis}
\label{sec:data_analysis}
The primary goal of our data analysis is to obtain the spatially resolved distributions of the two galaxy physical properties: stellar mass and SFR. In this respect, we carried out a SED fitting procedure on subkpc scales using the publicly available \texttt{magphys} code \citep{daCunha2008}. The methodo\-logy that we adopt to perform the data analysis has been developed and already used in previous works for a sample of local star forming galaxies with a grand-design spiral structure \citep{Enia2020, Morselli2020}. 
Our aim is to ensure a fair comparison with their results for unbarred galaxies, as they have consistent global properties with respect to ours. 

In \cite{Enia2020}, the procedure is summarised in three main steps: (i) degradation of the resolution of each image to the worst point spread function (PSF) that is available among all the photometric bands (i.e. that of the $350\,\upmu$m image taken with the SPIRE instrument, characterised by a PSF FWHM of \SI{24}{\arcsec}); (ii) the setting up of a grid of square cells of fixed size from which we measure the flux at different wavelengths; and (iii) derivation of the physical properties for individual apertures by performing SED fitting to the available photometric data. We analyse every selected source considering cells of $\SI{8}{\arcsec} \times\, \SI{8}{\arcsec}$ (corresponding to physical scales varying from $0.19\,\mathrm{kpc}$ to $0.77\,\mathrm{kpc}$, as reported in Table \ref{tab:properties}), which is the pixel scale of SPIRE $350\,\upmu$m maps. This choice is justified by the fact that we need a good balance between resolution and the number of available photometric bands in order to properly sample the SED ---including the dust emission curve--- and to avoid an excessively low resolution, which would prevent us from investigating the bar region. Specifically, it is necessary to sample the bar radius of each galaxy through at least three resolution elements (cells of $\SI{8}{\arcsec}$ each) in order to obtain accurate information about this morphological structure. At this point, we highlight the fact that, throughout the paper, we refer to the approximately annular region around
the very central nucleus that extends up to the radius of the bar as the `region hosting the bar'.

\subsection{Pre-processing and PSF degradation}
The preliminary steps that are performed on each photometric observation are those related to the preprocessing and PSF homogenisation phases, which start with the removal of foreground stars by masking contaminating sources through an automated procedure embedded in \texttt{CAAPR}. Subsequently, a background estimation and subtraction are performed on each of the star-subtracted maps following the guidelines provided by \cite{Clark2018}. Finally, the resolution of all the stars and background-subtracted maps is degraded to the PSF of SPIRE 350, exploiting the kernels of \cite{Aniano2011}.

\subsection{Photometry and SED fitting}
In order to perform the pixel-by-pixel SED fitting, it is necessary to measure the flux at each photometric band within the apertures. To this purpose, we generated a grid using a customisable process, through which the user can choose the cell size ($\SI{8}{\arcsec} \times\,\SI{8}{\arcsec}$, in our case) and the radius of the circle centred on the galaxy centre within which the grid is built. 
The fluxes at the various wavelengths are then measured in every cell using the \texttt{photutils} v0.6 Python package \citep{Bradley2019}. Those at wavelengths shorter than 10~$\upmu$m are also corrected for Galactic extinction based on values in the IRSA Galactic Dust Reddening and Extinction Service thanks to the in-built module in \texttt{CAAPR} \citep{Schlafly2011}. Subsequently, to give an estimate of the flux errors, we used the error maps in the DustPedia archive, which are typically available for the Spitzer bands and for the FIR photometry.
In cases where no error map is accessible (i.e. the SDSS maps), the uncertainty for a given flux is computed as the quadrature sum of the calibration noise and the root mean square (rms) of the maps, which is similar to the approach of the DustPedia collaboration \citep{Clark2018}. 
The outcome of this step of the data analysis is a photometric catalogue that contains the flux in the available photometric bands from FUV to FIR ---with the corresponding uncertainties--- along with the associated astrometric positions of the cell centres. It is worth mentioning that, for each cell, photometric bands that are characterised by a signal-to-noise ratio (S/N) of lower than 3 are rejected. Additionally, if a specific aperture exhibits more than ten rejected photometric bands due to an excessively low S/N, it is excluded from the subsequent analysis.

As mentioned above, we performed the SED fitting procedure using the \texttt{magphys} code. Through a Bayesian approach, it models the whole UV-to-FIR emission, assuming a balance bet\-ween the energy emitted at UV, optical, and NIR wavelengths that is absorbed by dust and the emission re-emitted in the FIR. In order to fit the observed multi-wavelength photometry, \texttt{magphys} relies on the \cite{Bruzual2003} stellar population synthesis code, which is used to compute the light produced by stars in galaxies. This code predicts the spectral evolution of stellar populations for varying star formation histories, which are described by a continuous model $\rm SFR(t) \propto exp^{-\upgamma t}$ with superimposed random bursts, following \cite{Kauffmann2003}. Galaxies are assumed to produce stars according to this law from a certain formation time t$_{\rm form}$, ranging between $0.1$ and $13.5\,\mathrm{ Gyr}$, to the present. The typical values of the star formation timescale parameter $\upgamma$ are approximately uniformly distributed over the interval from $0$ to $0.6\,\mathrm{Gyr}^{-1}$, dropping exponentially around $\upgamma = 1\,\mathrm{Gyr}^{-1}$. The models also include metallicities ranging from 0.2 to $2\,Z_{\odot}$. To characterise the attenuation of starlight by dust, the two-component model proposed by \cite{charlot2000} is employed \citep{daCunha2008}. This accounts for the luminosity absorbed and re-emitted by dust in stellar birth clouds (i.e. giant molecular clouds) and in the diffuse ISM in galaxies. The infrared emission from molecular clouds is characterised by the sum of three components: a component of polycyclic aromatic hydrocarbons (PAHs); a mid-infrared continuum associated with the emission from hot grains at temperatures of between 130 and 250 K; and a component of grains in thermal equilibrium with an adjustable temperature in the range of 30 -- 60 K. For the diffuse ISM emission, a component of cold grains with temperatures of 15 -- 25 K is also included.

Following \cite{SmithHayward2018}, the previously measured fluxes (in all the grid apertures) are given as an input to \texttt{magphys} for SED fitting in cells of $\SI{8}{\arcsec}$  per side. The criterion that is adopted to establish whether or not a fit has to be accepted or rejected is based on a $\chi^2$ cut \citep{Hayward2015, SmithHayward2018}, where such a threshold value is fixed according to the number of photometric bands that are available for the selected objects. In particular, as the galaxies of interest in this work are observed in a number of bands that can vary from 18 to 23, we adopt a conservative threshold value of 25 for the minimum $\chi^2$, remaining consistent with the choice made in \cite{Enia2020}. 

\subsection{Stellar mass and SFR estimates}
\label{subsec:mass_sfr_estimates}
Finally, the output of this spatially resolved SED fitting process, performed by the \texttt{magphys} code, consists of a wide range of physical properties. Specifically, we are interested in stellar mass ($M_{\star}$) and SFR, which are the two quantities that play a fundamental role in the characterisation of the MS relation.

Stellar masses are taken from the \texttt{magphys} output, while SFR estimates are obtained by summing the unobscured component (SFR$_{\rm UV}$) and the component that is obscured by dust absorption and then re-emitted in the IR (SFR$_{\rm IR}$). This empirical approach takes into account the contribution of two extreme stellar populations: the UV emission from young stars and the obscured young stellar component that is still hidden in the dusty molecular clouds. The SFR$_{\rm UV}$ contribution is estimated using the relation of \cite{Kennicutt1998}:
\begin{equation}
\label{eq:SFR_UV}
\mathrm{SFR}_{\rm UV} = 0.88 \cdot 10^{-28} L_{\rm UV,\nu}    
,\end{equation}
where $L_{\rm UV,\nu}$ is the luminosity at $150\,\mathrm{nm}$ expressed in erg s$^{-1}$ Hz$^{-1}$ and taken from the SED fit. The SFR$_{\rm IR}$ component is also derived from the relation of \cite{Kennicutt1998}:
\begin{equation}
\label{eq:SFR_IR}
\mathrm{SFR}_{\rm IR} = 2.64 \cdot 10^{-44} L_{\rm IR}    
,\end{equation}
where $L_{\rm IR}$, given in erg s$^{-1}$ by the SED fit, is the luminosity integrated between $8\,\mu$m and $1000\,\mu$m (rest-frame), which is a measurement of the energy that is absorbed by dust and re-emitted at MIR and FIR wavelengths. In both cases, the relations have been rescaled in accordance with a \cite{Chabrier_2003} IMF.

The SFR values are derived from the above empirical relations in order to reduce their dependence on the grid of \texttt{magphys} templates with respect to the ones directly obtained from the SED fit, which are, indeed, discretised and more dependent on degenerate parameters, such as age, metallicity, and extinction. This choice is also supported by the findings of \cite{Enia2020} regarding normal galaxies: the SFR computed as the sum of Equations \ref{eq:SFR_UV} and \ref{eq:SFR_IR} is consistent with the SFR given by \texttt{magphys} (see Appendix \ref{app:comparison}).


\begin{figure*}
    \centering
    \includegraphics[width=0.95\linewidth]{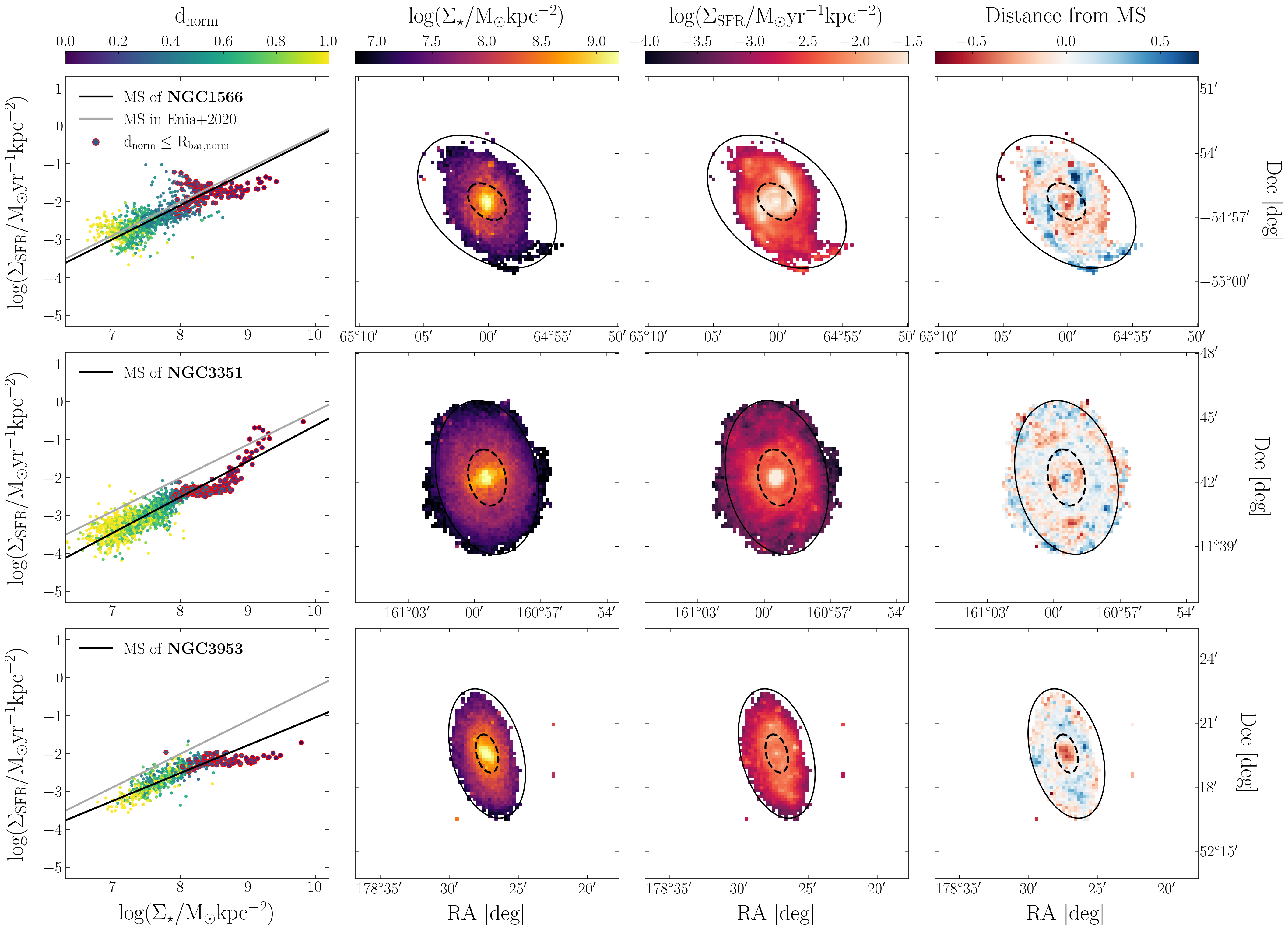}
    \caption{Results for NGC~1566 (first row), NGC~3351 (second row), and NGC~3953 (third row). For each galaxy, from left to right,  we show: (i) how the galaxy cells populate the $\log \Sigma_{\star}$ -- $\log \Sigma_{\rm SFR}$ plane with the corresponding best-fit MS (black line), as well as the MS found by \cite{Enia2020} for unbarred galaxies (grey line), where each point is colour coded according to the d$_{\rm norm}$ values and the ones sampling the bar-hosting region (d$_{\rm norm}$ $ \leq $ R$_{\rm bar,norm}$) are circled in red; (ii) the spatially resolved distribution of the stellar mass surface density; (iii) the spatially resolved distribution of the SFR surface density; and (iv) the map representing the distance of each cell from the best-fit MS. In the last three columns, we display the 25th level isophote of semi-major axis equal to R$_{25}$ with a solid black line and the rescaled ellipse enclosing the bar with a dashed black line.}
    \label{fig:panel_1}
\end{figure*}

\section{Results}
\label{sec:results}
In this section, we outline the key findings of our analysis,
starting with the presentation of the spatially resolved distributions of the stellar mass surface density $\Sigma_{\star}$ and of the SFR surface density $\Sigma_{\rm SFR}$. The second and
third columns of Figs. \ref{fig:panel_1} and \ref{fig:panel_2} display maps of $\Sigma_{\star}$ and $\Sigma_{\rm SFR}$, respectively. All these plots also include: a solid black ellipse of semi-major axis equal to R$_{25}$, representing the 25th level isophote of each galaxy, and a dashed black ellipse with the same axis ratio as the 25th isophote but rescaled in accordance with the bar radius. 

We then show the spatially resolved MS relations of all the sources, which are collected in the first column of Figs.\,\ref{fig:panel_1}--\ref{fig:panel_2}, as well as our analysis of them. As it can be seen from these plots, the distribution of points (each one representing the values of $\Sigma_{\star}$ and $\Sigma_{\rm SFR}$ in a certain grid cell) is colour coded according to the normalised distance of the aperture from the galaxy centre: d$_{\rm norm}$. The normalisation of this radial distance is performed with respect to the 25th level isophote. Specifically, the distance of each cell from the centre is divided by the radial distance of the point lying on the ellipse of the 25th isophote along the direction identified by that cell. 
The points that are circled in red correspond to the square apertures lying within the dashed black ellipse, meaning that they are sampling the bar region. In particular, once the bar radius is normalised with the aforementioned procedure (R$_{\rm bar,norm}$), we identify these points as the ones characterised by d$_{\rm norm}$ $ \leq $ R$_{\rm bar,norm}$. 

For each galaxy, the best-fit MS is displayed as a black line, while the MS obtained in the work of \cite{Enia2020} for a sample of unbarred grand-design spiral galaxies from the DustPedia archive is reported as a grey line. The latter is given by the following expression:
\begin{equation}
\label{eq:EniaMS}
\log\Sigma_{\rm SFR} = 0.88 \cdot \log\Sigma_{\star} - 9.05   
.\end{equation}

In the end, the fourth column of Figs.\,\ref{fig:panel_1}--\ref{fig:panel_2} collects the spatially resolved maps that are colour coded as a function of the distance from the best-fit MS of each galaxy. Also in this case, the solid and dashed black ellipses mark the 25th isophote and the bar region, analogously to what is also represented in $\Sigma_{\star}$ and $\Sigma_{\rm SFR}$ maps.

\subsection{Spatially resolved distributions of \texorpdfstring{$M_{\star}$}{star} and SFR}

\begin{figure*}
    \centering
    \includegraphics[width=0.95\linewidth]{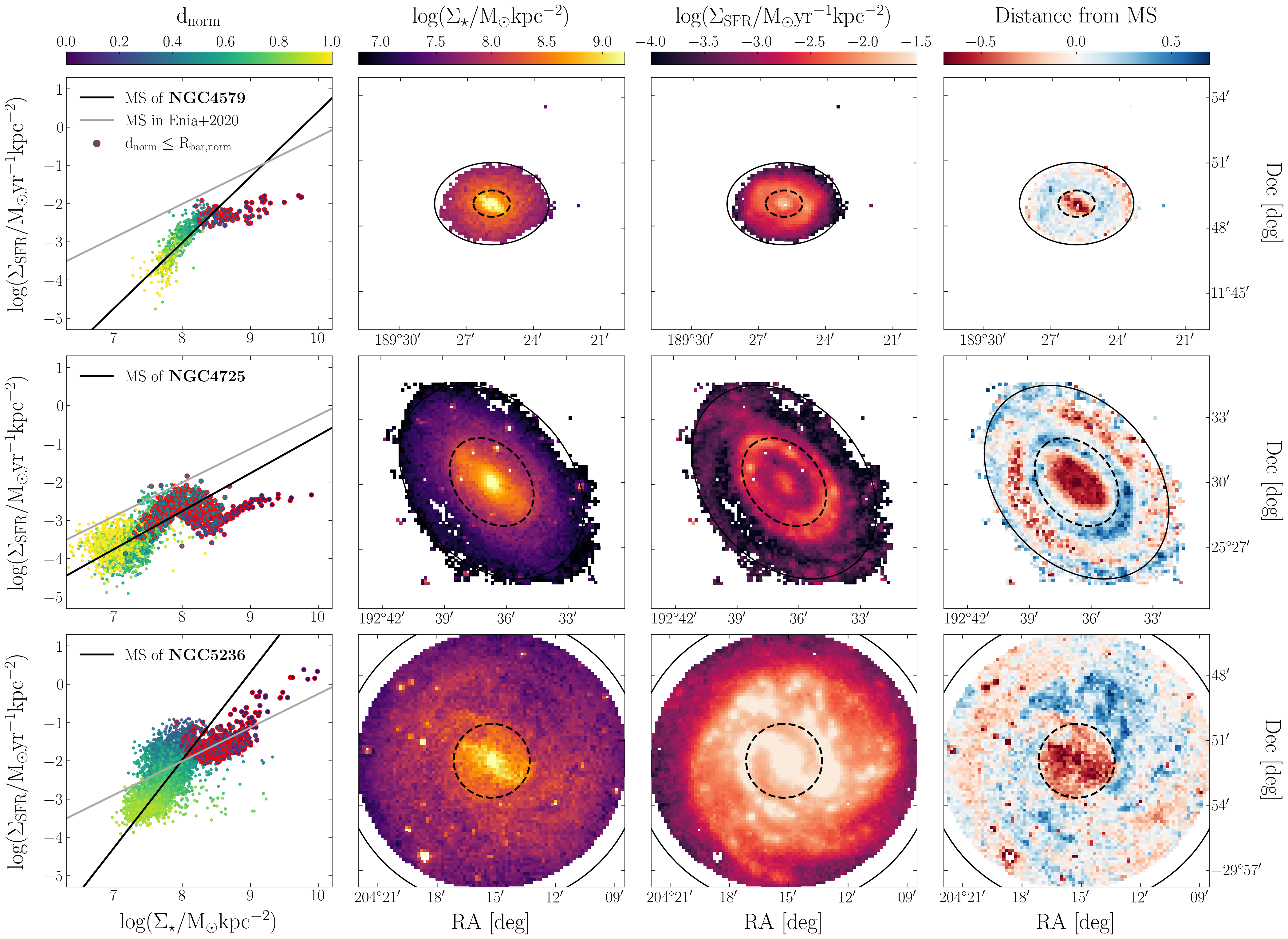}
    \caption{Same as Fig. \ref{fig:panel_1}, but for NGC~4579, NGC~4725, and NGC~5236.}
    \label{fig:panel_2}
\end{figure*}

As already mentioned in Sect.\,\ref{subsec:mass_sfr_estimates}, the stellar mass values of the single cells are obtained from the \texttt{magphys} output. Then, it is straightforward to get the corresponding surface density by dividing the stellar mass of each grid element by its area. 
In all the galaxies of our sample, the stellar mass surface density (second column plots of Figs.\,\ref{fig:panel_1}--\ref{fig:panel_2}) appears to smoothly increase, going from the galaxy outskirts towards the inner regions, as expected for the typical exponential growth of disc galaxies. Additionally, an almost spheroidal structure in the centre is evident in every galaxy and the stellar bar appears to be grafted onto it.

Regarding the distributions of $\Sigma_{\rm SFR}$ (third column plots of Figs.\,\ref{fig:panel_1}--\ref{fig:panel_2}) obtained by dividing the sum of SFR$_{\rm UV}$ and SFR$_{\rm IR}$ by each cell area, we note that the spiral arms are well traced in most of the galaxies, and are particularly well traced in NGC~1566 and NGC~5236. Relatively high values of star formation are typically evident in several peaks along the spiral arms and in some of the very central regions (e.g. the nucleus of NGC~3351). The only exception is NGC~3953, which shows an overall smoother distribution of SFR density. In addition, it is possible to identify a nearly annular and more quiescent structure (lower values of $\Sigma_{\rm SFR}$) that coincides with the region that hosts the bar (the one enclosed by the dashed black ellipse). This is obvious in NGC~4725, but is also detectable in NGC~3351 and NGC~4579. The outermost regions of all the selected galaxies exhibit the lowest star formation activity. The most striking difference between the spatially resolved distributions of $\Sigma_{\star}$ and $\Sigma_{\rm SFR}$ is that stellar mass clearly traces the presence of the bar, while it is almost never visible in SFR maps.

\subsection{Spatially resolved MS relation}
The ultimate goal of this study is to probe the MS relation on subkpc scales for our sample of barred galaxies. Accordingly, after exploring the spatial distribution of the two physical properties of interest here, it is necessary to relate them and analyse the distribution in the $\log \Sigma_{\star}$ -- $\log \Sigma_{\rm SFR}$ plane (first column plots of Figs.\,\ref{fig:panel_1}--\ref{fig:panel_2}). 

In general, the stellar mass density increases towards the centre of the galaxies, as does the density of the SFR, in agreement with the expected log-linear relation. Nevertheless, the presence of a common decreasing feature arises when looking at the overall trend of the MS in the central part of all these barred galaxies. Indeed, a ‘bending’ of the MS relation is observed in correspondence with the points sampling the region where the bar is located, which are the ones circled in red.
In most cases, the values of $\Sigma_{\rm SFR}$ clearly tend to increase again at the high $\Sigma_{\star}$ end.  

For each galaxy, the functional form of the spatially resolved MS relation is obtained by performing a linear fit ---through the orthogonal distance regression function implemented in the \texttt{scipy} package of Python--- of the values in the $\log \Sigma_{\star}$ -- $\log \Sigma_{\rm SFR}$ plane derived for all the cells within that galaxy. In Table \ref{tab:MSslopes_dustpedia}, we display slopes $m$, zero points $q$ (with their corresponding uncertainties), and scatter $\sigma$ of the fitted MS relations for all six barred galaxies. The latter is computed as the standard deviation of the residuals along the y-axis with respect to the best-fit MS.
\begin{table}
    \centering
    \caption{Parameters of the best-fit MS of each galaxy in the sample.}\label{tab:MSslopes_dustpedia}
    \begin{tabular}{lccc}
        \hline
        \hline
        Galaxy Name & m$\pm\Delta$m & q$\pm\Delta$q & $\sigma$\\
        \hline
        \noalign{\vskip 1pt}
        NGC~1566   &  $0.89 \pm 0.09$  & $-\phantom{0}9.23 \pm 0.71$ & 0.31\\
        NGC~3351   &  $0.94 \pm 0.07$  & $-10.06 \pm 0.53$ & 0.23\\
        NGC~3953   &  $0.73 \pm 0.10$  & $-\phantom{0}8.38 \pm 0.80$ & 0.20\\
        NGC~4579   &  $1.69 \pm 0.23$  & $-16.54 \pm 1.85$ & 0.36\\
        NGC~4725   &  $0.99 \pm 0.05$  & $-10.69 \pm 0.39$ & 0.39\\
        NGC~5236   &  $2.29 \pm 0.12$  & $-20.31 \pm 0.92$ & 0.56\\
        \hline     
    \end{tabular}
    \tablefoot{Slopes (m), zero points (q), associated uncertainties ($\Delta$m and $\Delta$q), and scatter ($\sigma$) of the fitted MS relations for the six barred galaxies.}
\end{table}
As already mentioned, the best-fit lines are shown as black lines in the first column plots of Figs.\,\ref{fig:panel_1}--\ref{fig:panel_2}. 
By comparing the resolved MS relation of each of our barred galaxies with the one obtained for normal spirals (Eq. \ref{eq:EniaMS}, grey line; \cite{Enia2020}), we observe that the distribution of points typically lies below the relation found for unbarred galaxies, which is consistent with a systematically lower SFR at fixed stellar mass. This is valid for NGC~3351, NGC~3953, NGC~4579, and NGC~4725, while NGC~1566 and NGC~5236 show higher values of $\Sigma_{\rm SFR}$, as expected from the presence of highly star-forming spiral arms and nuclear regions. Furthermore, the common trend of MS bending in the bar-hosting regions appears to be anticorrelated with the best-fit relations (black lines), although this effect is more striking in some galaxies than in others. 

While the properties discussed so far capture the overall trend of the spatially resolved MS relation of barred galaxies, there are still several galaxy-to-galaxy variations that need to be highlighted. The best-fit MS of NGC~1566 is almost overlapping with the MS of normal spirals and they are consistent within the uncertainties. In NGC~3351, NGC~3953, and NGC~4725, we note that the slopes of the fitted MS relations are similar to those of unbarred spirals, while the zero points have smaller values. On the other hand, the fitted MS of both NGC~4579 and NGC~5236 shows a steeper slope than that found for unbarred galaxies, meaning that the increase in SFR towards the galaxy centre is more rapid. However, while NGC~4579 presents values of $\Sigma_{\rm SFR}$ that are globally lower than the typical trend of normal galaxies, NGC~5236 stands out as a highly star-forming source. 

Focusing on the inner regions of the galaxies in order to better investigate the impact of the bar structure, we can point out that NGC~4579, NGC~4725, and NGC~5236 (Fig. \ref{fig:panel_2}) are the galaxies with the most evident feature of anti-correlation. Indeed, if the points associated with the region that hosts the stellar bar turn out to be arranged in a direction approximately orthogonal to the fitted MS, then the $\Sigma_{\rm SFR}$ decreases with increasing mass density. Moreover, going towards the very centre of these three galaxies, it is remarkable that the SFR density reverts to having an increasing trend as the stellar mass density grows; however, these very central points still appear to lie below the best-fit MS of each galaxy. This implies that the nuclear regions are less star-forming than what would be predicted from $\Sigma_{\star}$ based on their own MS. The other three galaxies in our sample, namely NGC~1566, NGC~3351, and NGC~3953 (Fig. \ref{fig:panel_1}), exhibit a smoother decline in $\Sigma_{\rm SFR}$ in the bar region. In particular, the MS of NGC~1566 is also more scattered than the others, reflecting the wider range of $\Sigma_{\rm SFR}$ values that mostly characterise the winding spiral arms. In addition, as opposed to the other galaxies, in NGC~3351 we note that the closest points to the galaxy centre lie above the best-fit relation, which suggests a more active star-forming region than what would be foreseen by the MS, given their stellar mass density.

\subsection{Distance from MS}

The panels in the last column of Figs. \ref{fig:panel_1}--\ref{fig:panel_2} provide some information about what is happening in each of these galaxies: the red cells correspond to regions that are more weakly star forming than what the fitted MS would predict, while the blue ones are associated with particularly active, starbursting regions. The central parts (including both the nuclear and the bar hosting regions) of most of the selected objects appear to be red, with suppressed or lower levels of star-formation activity compared to the global galaxy MS relation, or perhaps even quenched star formation. On the other hand, their spiral arms show a tendency to be active in star formation. On the contrary, NGC~3351 exhibits a star-forming nucleus with a red (quiescent) annular region around it.


\section{Discussion}
\label{sec:discussion}

The leading outcome of our work is the discovery of an anti-correlating feature in the spatially resolved MS relations associated with an almost annular region that is hosting the bar. This pattern suggests that the typical MS bending observed at the high $\Sigma_{\star}$ end, both in the spatially resolved relation \citep[e.g.][]{Abdurro'uf2017} and in the integrated one \citep[e.g.][]{Popesso2019a}, could be due to the presence of a stellar bar and not only the bulge, as usually suggested \citep[e.g.][]{Morselli2019}. However, the precise drivers of this bending of the MS remain unknown. For this reason, it could be crucial to investigate this correlation across different morphological structures \citep[e.g.][]{Pessa2022}. The fact that the bar-hosting region is typically more weakly star forming and is probably on the way to being quenched (it shows lower values of SFR with respect to those predicted by the scaling relation characterising the disc) supports the inside-out quenching scenario, which is supposed to play an important role in driving galaxies below the MS \citep[e.g.][]{Tacchella2015, Tacchella2018, Abdurro'uf2018}. This observed feature could also suggest co-evolution of the bar structure and the inner bulge: the bar favours star formation quenching and thus induces bulge growth.

The quenching track in the spatially resolved MS can be more or less prominent: it is characterised by different slopes depending on the host galaxy (Figs.\,\ref{fig:panel_1}--\ref{fig:panel_2}). This variability can be interpreted as a consequence of the different evolutionary stages in which the stellar bar is observed. More recently formed structures are probably still funnelling cold gas towards the inner regions of their hosts, resulting in higher SFRs and a less obvious anti-correlation feature (e.g. NGC~1566). On the other hand, during the late evolutionary stages of stellar bars, the central regions of their host galaxies are expected to be nearly devoid of fuel for further star formation as a consequence of the previous rapid consumption of gas. This could potentially result in a more pronounced quenching track (e.g. NGC~4725), thereby also favouring bulge growth.

As mentioned in Sect. \ref{subsec:environment}, the environment in which galaxies reside is believed to influence their properties and evolution. The galaxies in our sample are located in quite different environments: field, loose groups, and clusters. Among the various external processes, ram pressure stripping or gravitational interactions may have influenced the outskirts of some galaxies, and  so we cannot completely exclude environmental effects. However, the presence of the anti-correlation track in the bar region seems to be common to all six objects, albeit with different levels of prominence. Therefore, we conclude that the environment is not the main driver of the feature found in our results; at most, it could be responsible for scatter or second-order variations.

In the majority of the galaxies of our sample, the nuclear regions are less star forming than the spiral arms, except for the case of NGC~3351, which appears to be still active in forming stars in the very centre. A more obvious dual behaviour of the inner regions was pointed out by \cite{Enia2020} for their unbarred galaxies; approximately half of their sample exhibits higher values of $\Sigma_{\rm SFR}$ in the nucleus, and the other half are more quenched. Although we do not have sufficient statistics to infer the actual difference, the fact that our barred galaxies are usually more quiescent in the central regions suggests a possible effect of the  presence of the bar. Indeed, this morphological structure is thought to be able to funnel gas both inwards and outwards and this can ignite starburst episodes, thus producing a rapid consumption of fuel for star formation and the onset of a quenching phase \citep[e.g.][]{FraserMcKelvie2020a, Geron2021}.

\begin{figure}[t]
    \centering
    \includegraphics[width=1.0\linewidth]{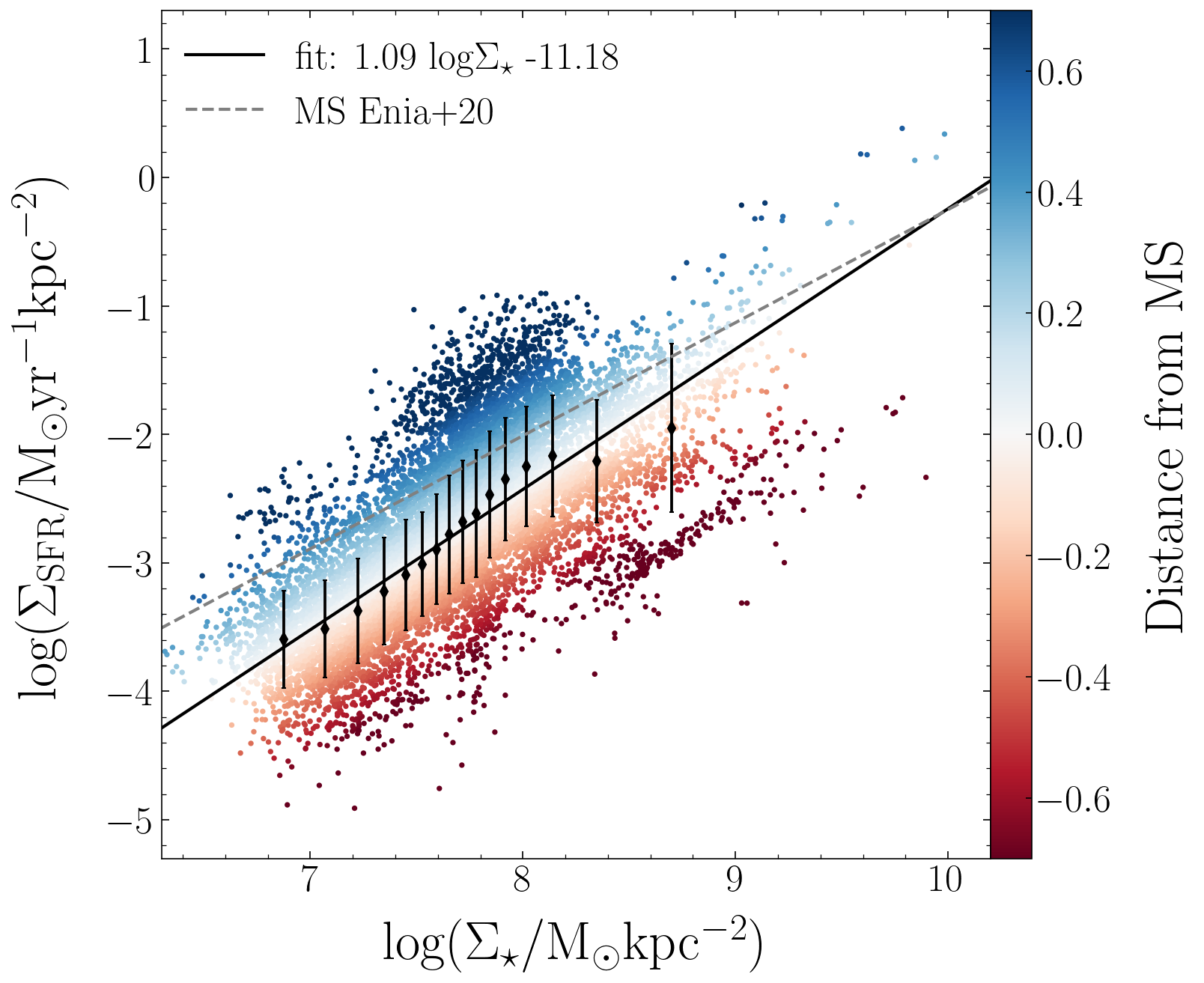}
    \caption{Overall MS relation. Each point represents the values of the stellar mass and SFR surface density of an accepted cell in one of the six galaxies of our sample. These points are colour coded as a function of their distance from the MS (black solid line), which is derived from the linear fit of the black points, which are marked with thin diamonds. The error bars correspond to the standard deviation of the data points in each bin. The grey dashed line represents the best-fit MS obtained in \cite{Enia2020} for a sample of unbarred spiral galaxies.}
    \label{fig:total_MS}
\end{figure}

Similar results pointing to lower levels of star formation in the bar region were also obtained by \cite{Pessa2021} based on spatially resolved SFR measurements from optical emission line maps, confirming the idea of `star formation deserts' proposed by \cite{James2018}.
In an attempt to capture a possible common trend in the star formation activity of barred galaxies, we computed the MS relation in the $\log \Sigma_{\star}$ -- $\log \Sigma_{\rm SFR}$ plane, considering the accepted grid cells of all six galaxies (see Fig.\,\ref{fig:total_MS}). This is also done to provide a fair comparison with the final MS relation found in \cite{Enia2020} (Eq. \ref{eq:EniaMS}). Simply looking at the distribution of points, it is evident that different galaxies populate distinct portions of the plane, resulting in a quite broad relation, where the typical anti-correlation feature is barely discernible. This suggests that there is significant variability from one galaxy to another, as is highlighted in the presentation of our results (see Sect.\,\ref{sec:results}). To find the overall best-fit MS, we subdivided the mass-density interval into 16 bins. Within each bin, we computed the median log$\Sigma_{\star}$ and log$\Sigma_{\rm SFR}$, while the error is taken as the standard deviation of the points inside the bin. In addition, we fitted our binned data points (black diamonds markers with associated error bars in Fig.\,\ref{fig:total_MS}) with a log-linear relation using \texttt{emcee} \citep{Foreman-Mackey2013} and obtain: 
\begin{equation}
\label{eq:totMS_bar}
\log \Sigma_{\rm SFR} = m \cdot \log\Sigma_{\star} + q   
,\end{equation}
where $ m = 1.09^{+0.24}_{-0.25}$ and $ q = - 11.18^{+1.90}_{-1.82}$ dex. The global scatter of this final fit is $\sigma = 0.48$ dex. The values for the slope and the zero point of the best-fit relation are consistent ---within the uncertainties--- with those found by \cite{Enia2020}, reported in Eq. \ref{eq:EniaMS}. However, our relation is more scattered; indeed they found the scatter of the MS of unbarred galaxies to be $\sim 0.27$ dex. This implies that there is greater variability in the properties of barred galaxies compared to those of unbarred objects, although these findings are based on limited statistics. Moreover, in Fig. \ref{fig:total_MS}, it is evident that the majority of data points are distributed below the MS of unbarred galaxies (grey dashed line). Hence, barred galaxies are generally less star forming than their unbarred counterparts. This is also confirmed when looking at the integrated properties of the galaxies in our sample and comparing them with those in the work of \cite{Enia2020}. Within the same range of total stellar mass, that is, within $10^{10} - 10^{11}$ $ \si{M_\odot}$, the SFRs of unbarred galaxies are between 1 and 4 $ \si{M_\odot}$yr$^{-1}$, while barred galaxies show values of SFR well below 1 $ \si{M_\odot}$yr$^{-1}$ (with the only exception being NGC~5236; see Table \ref{tab:properties}).

\begin{figure*}
    \centering
    {\includegraphics[width=1.0\linewidth]{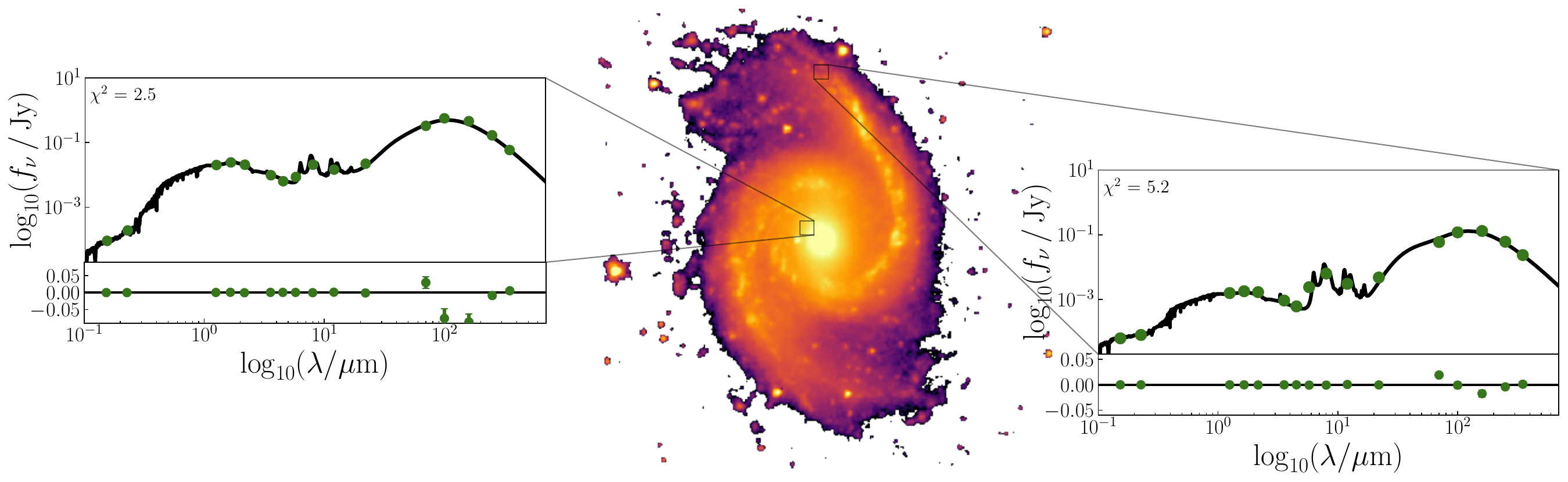}}  \quad
    {\includegraphics[width=1.01\linewidth]{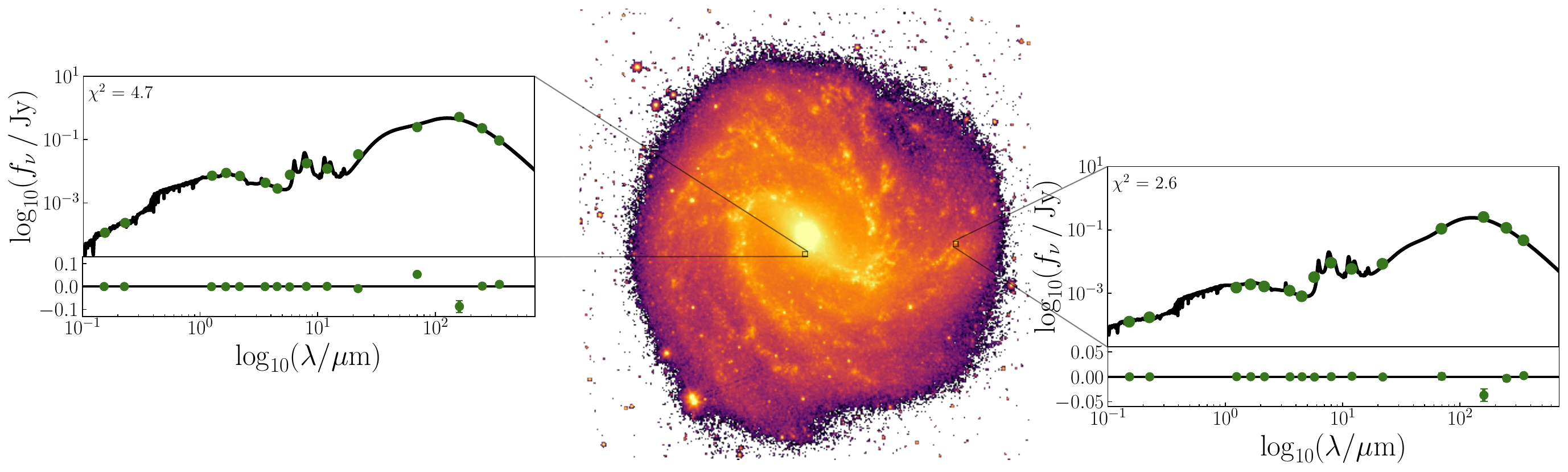}}
    \caption{Output of the \texttt{magphys} SED fitting performed in one of the central pixels and in one of the outermost pixels of NGC~1566 (upper panels) and of NGC~5236 (bottom panels). The green data points are the observed fluxes in the available bands and the $\chi^2$ of the fit (black line) is reported in the upper left corner of the plots. At the bottom of each subfigure, we also display the difference in dex between the observations and the best-fitting model with the associated error bars; when not visible they are smaller than the corresponding markers.}
    \label{fig:sed_agn_sf}
\end{figure*}

As anticipated in Sect.\,\ref{sec:data_sample}, two out of six galaxies (namely NGC~1566 and NGC~4579) are characterised by low AGN activity in their nuclear regions. As the \texttt{magphys} code does not account for a possible AGN contribution to the SED, we checked its output by visually inspecting the SED fitting results obtained for the central pixels, where fluxes might be dominated by AGN emission. In Fig.\,\ref{fig:sed_agn_sf} we show the outcome for one of the central pixels and for one of the outermost pixels of NGC~1566, one of the two galaxies hosting a LLAGN, and of NGC~5236, which has no nuclear activity, for comparison. 
According to our results, the SED fit in most of the central pixels of NGC~1566 does not exhibit any peculiar feature that can be attributed to the influence of an AGN (as clearly visible in the example shown in Fig.\,\ref{fig:sed_agn_sf}). The only exception is for the pixel sampling the very centre of the galaxy, where the AGN contribution manifests as a partial suppression of the PAHs features. Hence, we ultimately excluded the very central pixel of both NGC~1566 and NGC~4579 from our analysis, to avoid any potential contamination. However, it is worth highlighting that this choice does not affect our results, as the nuclear region still remains properly sampled.
Furthermore, looking at the SED fit in the pixels that are just slightly offset with respect to the very centre, we notice that there is no obvious residual in the $\sim 3$ -- $8\,\mu$m range (covered by WISE and the IRAC camera on board \textit{Spitzer}), where the dusty torus emission of the AGN would have appeared \citep[e.g.][]{Fritz2006}. The absence of significant excess in this wavelength range with respect to the modelled continuum flux indicates that the resulting star formation is sufficient to explain the observed SED. This is also confirmed by the fact that, comparing the SED fit in one of the central pixels of NGC~1566 with that of NGC~5236, there are no significant differences in the shape of the best-fit curve (left panels of Fig.\,\ref{fig:sed_agn_sf}). The conclusions listed above are valid for both NGC~1566 (upper panels of Fig.\,\ref{fig:sed_agn_sf}) and NGC~4579 and support our initial decision to include LLAGN-hosting galaxies in our sample, assuming that the presence of a LLAGN would not significantly affect our SED modelling. Concerning NGC~4579, we have to underline the fact that, although the LLAGN seems not to have a relevant impact on the SED fitting of the central pixels, the suppressed SFR detected in the region hosting the bar may also be due to a radio jet cocoon and radio-jet-driven outflows into the multi-phase ISM, as found by \cite{Ogle2023}. Additionally, comparing the SED fit in the innermost subkpc region of our galaxies with that obtained in one of the outer pixels, we note that they have a similar shape, supporting the idea that the LLAGN makes a negligible contribution to the galaxy emission.   

Nevertheless, many controversial results emerged from previous studies when exploring the connection between the bar structure and the presence of a potential AGN \citep[see][for reviews]{Combes2003, Combes2006, Jogee2006}. More recently, while \cite{Cisternas2013}, \cite{Goulding2017}, and \cite{Silva-Lima2022} claimed that bars are not among the main mechanisms driving AGN feeding, \cite{Cisternas2015} and \cite{Alonso2018} revealed that there could be an interplay between bars and AGN, supporting the idea of bar-induced nuclear activity. Alternatively, \cite{Zee2023} suggested that AGN evolution can regulate the shape of a bar: these elongated structures can be destroyed by the growth of a central mass concentration or a black hole. 
Interpretation of these disparate results concerning the joint evolution of bars, AGN, and star formation is still a highly active area of research.


\section{Summary and conclusions}
\label{sec:conclusions}
In this work, we study the spatially resolved MS relation for six almost face-on barred spiral galaxies in the local Universe, with the aim of unveiling the relevance of possible bar-driven quenching effects. We took advantage of the publicly available DustPedia archive to collect photometric information on UV to FIR wavelengths for the entire sample. We performed a panchromatic SED fitting procedure on subkpc scales using the \texttt{magphys} code. We produced maps of stellar mass and SFR surface density and compared them in the $\log \Sigma_{\star}$ -- $\log \Sigma_{\rm SFR}$ plane.

Our key findings can be summarised as follows:
\begin{itemize}
    \item All the selected galaxies show a peculiar feature in the resolved MS relation: the bar-hosting region tends to be offset below the best-fit MS, whose slope is largely determined by the dominant disc structure. Indeed, a slight anti-correlation in correspondence with the annular region that hosts the bar is clearly visible in the MS of every galaxy and this suggests a suppressed SFR, which could be associated with the onset of a quiescent phase. We argue that stellar bars are mostly responsible for depleting gas reservoirs, thus halting star formation in the central regions. This supports the inside-out quenching scenario and the idea that bars may affect the typically observed bending of the MS at high $\Sigma_{\star}$. 
    \item We confirm a large spread among barred galaxies in the $\log \Sigma_{\star}$ -- $\log \Sigma_{\rm SFR}$ plane. Indeed, we infer a large scatter (0.48 dex) around the best-fit relation with a slope of $1.09$ and a zero point of $-11.18$, suggesting that galaxy-to-galaxy variations are in place.         
    \item The innermost central regions typically associated with a galactic bulge structure are characterised by lower  SFR values than those predicted by the best-fit MS, with the exception of NGC~3351. 
    \item Comparing our results with those obtained for a sample of normal spiral galaxies, we note that barred objects are typically less actively star forming than their unbarred counterparts, even though the slope and the zero point of our total MS are consistent within the uncertainties.
\end{itemize}

In future follow-up works, it would be useful to improve the statistics by selecting a larger sample of barred galaxies in order to further investigate the role of bar quenching and to elucidate a possible causal relationship between bar formation and the central ‘star-formation desert’. In particular, it may be interesting to include barred and unbarred galaxies hosting AGN to investigate whether or not a systematic difference is found compared to galaxies with no strong AGN activity. Moreover, high-resolution simulations (e.g. Illustris TNG) might be crucial in this kind of study in order to test the reproducibility of the observed quenching trend and to reconstruct the history of bar formation and the evolution of SFR. In this way, one should also be able to better constrain the actual impact of stellar bars on the suppression of SFR by quantifying this quenching mechanism over time from a cosmological perspective.



\begin{acknowledgements}
We thank the anonymous referee for the useful comments and suggestions that improved the overall work quality.

L.S. acknowledges the financial support from the PhD grant funded on PNRR Funds Notice No. 3264 28-12-2021 PNRR M4C2 Reference IR0000034 STILES Investment 3.1 CUP C33C22000640006.
G.R. and P.C. are supported  by the European Union – NextGenerationEU RFF M4C2 1.1 PRIN 2022 project 2022ZSL4BL INSIGHT.

A.E., P.C., G.R., and E.M.C. acknowledge the support from MIUR grant PRIN 2017 20173ML3WW-001 and Padua University grants DOR 2021-2023. E.M.C. is also funded by INAF through grant PRIN 2022 C53D23000850006. A.B. and G.R. acknowledge support from INAF under the Large Grant 2022 funding scheme (project ``MeerKAT and LOFAR Team up: a Unique Radio Window on Galaxy/AGN co-Evolution”). CG acknowledges the support from grant PRIN MIUR 2017 20173ML3WW-001 ``Opening the ALMA window on the cosmic evolution of gas, stars, and supermassive black holes'', and co-funding by the European Union – NextGenerationEU within PRIN 2022 project n.20229YBSAN ``Globular clusters in cosmological simulations and in lensed fields: from their birth to the present epoch”. V.C. acknowledges funding from the INAF Mini Grant 2022 program “Face-to-Face with the Local Universe: ISM’s Empowerment (LOCAL)”.

\end{acknowledgements}

%
%
\bibliographystyle{aa}
\bibliography{mybib}


\begin{appendix}
\section{SFR = SFR$_{\rm UV}$ + SFR$_{\rm IR}$ vs SFR$_{\rm MAGPHYS}$}
\label{app:comparison}
\begin{figure}[h]
    \centering
    \includegraphics[width=0.8\linewidth]{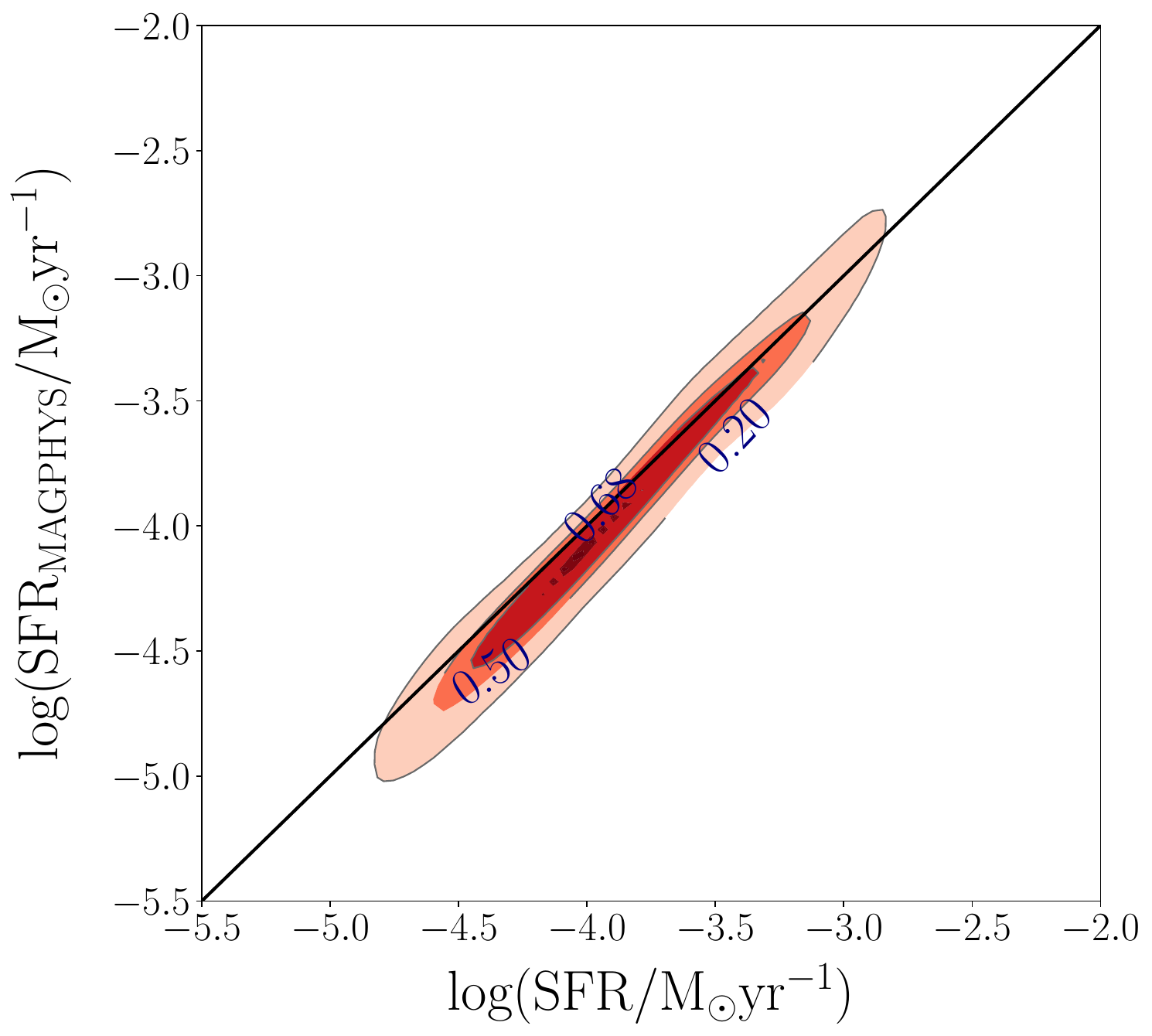}
    \caption{Comparison between the SFR values obtained from \texttt{magphys} 
    (y-axis) and the ones used in this work (x-axis), derived as the sum of SFR$_{\rm UV}$ (Eq. \ref{eq:SFR_UV}) and SFR$_{\rm IR}$ (Eq. \ref{eq:SFR_IR}). The red-scale contours are related to the distribution of points and three density levels are highlighted. The 1:1 relation is represented by the black solid line.}
    \label{fig:sfr_comparison}
\end{figure}

In this section, we present a comparison between the SFRs derived as the sum of the two empirical relations (Eqs. \ref{eq:SFR_UV}--\ref{eq:SFR_IR}) and the ones obtained from \texttt{magphys}.
Similarly to what has been done in the work of \cite{Enia2020}, we tested the consistency of the values of SFR producing the plot in Figure \ref{fig:sfr_comparison}. The scatter of the relation, represented by the red-scale contours, is given by the median absolute deviation and has a value of $\sim$ 0.15 dex, which is small. The fact that this tight distribution of points is compatible with a 1:1 relation (black line) ensures that the use of SFRs computed as the sum of Eqs. \ref{eq:SFR_UV} and \ref{eq:SFR_IR}, instead of those given by \texttt{magphys}, does not change the outcomes of this work.

\end{appendix}

\end{document}